\documentclass[12
pt]{article}
\usepackage{geometry}
\usepackage{float}
\usepackage{placeins}
\usepackage{subcaption}
\usepackage{makeidx}
\usepackage[T1]{fontenc}
\usepackage[dvipsnames,svgnames,table]{xcolor}
\usepackage{epstopdf}
\usepackage{epsfig}
\usepackage{textcomp}
\usepackage{hyperref}
\usepackage{amsmath}
\usepackage{amssymb}
\usepackage{multirow}
\usepackage{multicol}
\usepackage{booktabs}
\usepackage{lastpage}
\usepackage{hyperref} 
\usepackage{graphicx}
\usepackage{enumerate}
\usepackage{threeparttable}
\usepackage{float}
\usepackage{array}
\usepackage{cite}
\usepackage{color,soul}
\usepackage{listings}
\usepackage{tikz}
\makeatletter
\renewcommand\section{\@startsection{section}{1}{\z@}%
{-2.5ex \@plus -1ex \@minus -.2ex}%
{2.3ex \@plus.2ex}%
{\normalfont\large\bfseries}}
\renewcommand\subsection{\@startsection{subsection}{1}{\z@}%
{-2.5ex \@plus -1ex \@minus -.2ex}%
{2.3ex \@plus.2ex}%
{\small\bfseries}}

\begin{document}
\bigskip
\title{\textbf{Contribution of cosmic alphas to generation of albedo neutrons and gammas in lunar regolith: A computational study based on GEANT4 simulations}}
\medskip
\author{\small Harshala Gaonkar$^{1}$, A. Ilker Topuz$^{2}$}
\medskip
\date{\small 
$^1$Department of Physics, Sir Parashurambhau College, Savitribai Phule Pune University, Pune, Maharashtra, India 411030\\
$^2$Department of Physics, SRM University-AP, Amaravati 522240, India\\
email: ahmetllker.t@srmap.edu.in
}
\maketitle
\begin{abstract}
The lunar surface is constantly and directly exposed to Galactic Cosmic Rays (GCRs) due to the lack of the atmosphere and the magnetic field around it. The high-energetic alpha particles subsequently contribute to the generation of the secondary neutrons through either the spallation reaction or the low-energetic $(\alpha, n)$ process along with the cosmic protons. In this study, the contribution of the alpha particles to the production of the albedo neutrons as well as the albedo gamma rays in the lunar regolith is studied by using the Monte Carlo simulations based on the GEANT4 toolkit. The incident alpha particles are referenced from the PAMELA spectrometer over the interval between 0.14 and 52 GeV and executed as a probability-based primary particle source. The present simulations demonstrate the production of the albedo neutrons and gamma rays upon the process of cosmic alpha irradiation in the lunar regolith. The albedo neutron production is predominantly situated within the near-surface region of the lunar regolith and progressively diminishes with the increasing depth. The population of the secondary neutrons is primarily distributed over the low-energy region and reduces towards the high energies. These results imply that the cosmic alpha particles have also quantifiable contribution in the generation of the secondary radiation at the lunar surface together with the cosmic protons, and this study demonstrates a computational framework in the evaluation of the lunar albedo, the radiation transport, the surface composition, and the radiation environment for the further lunar exploration missions.
\end{abstract}
\textbf{\textit{Keywords: }} Cosmic alphas; PAMELA spectrometer; Albedo neutrons; Albedo gammas, Lunar regolith; GEANT4.
\section{Introduction}
Unlike earth, the Moon lacks in substantial atmosphere or global magnetic field for the effective shielding against the high energetic particles. Galactic Cosmic Rays (GCRs), originating from outer space, therefore interact directly with the lunar surface. GCRs consist predominantly of protons, followed by alpha particles and other heavy nuclei~\cite{simpson1982elemental,mewaldt1994galactic,blasi2013origin,tatischeff2021origin}. The alpha particles contribute around 10-15$\%$ of the total GCR flux after the protons and hence have a significant role in the generation of secondary particles such as the secondary neutrons via the spallation reaction or the low-energetic $(\alpha, n)$ process. When the energetic cosmic-ray particles penetrate the lunar regolith, they undergo electromagnetic and nuclear reactions with the constituent nuclei. These interactions further result in the nuclear cascades and lead to the production of the secondary particles, which include the neutrons and the gamma rays. A part of these secondary particles propagates towards the lunar surface and escapes, forming the lunar particle albedo. The characteristics of this albedo depends upon various factors such as type and energy distribution of primary incident particles, the elemental composition and density of lunar regolith, also the physical models used to illustrate particle-matter interactions \cite{zaman2022modeling,naito2023global,li2017simulation}.

The albedo neutrons play a significant role due to their ability to propagate through the lunar regolith without the continuous interactions. Their energy distribution and transport are influenced by the factors such as regolith composition, nuclear reactions, and scattering. The albedo neutrons further react with the other nuclei, resulting in the generation of secondary gamma radiation via the inelastic $(n, n'\gamma)$ scattering or the absorption $(n,\gamma)$ process. Consequently, the neutron and gamma ray emissions provide some valuable information about both the lunar radiation environment and the particle interactions occurring beneath the surface. This line of investigation builds on the previous work addressing the influence of hydrogen and particle-generation methodology on the lunar albedo neutrons \cite{topuz2024effect,topuz2022dome,ilker2023particle}. As the incident cosmic-ray spectrum is strongly energy dependent and ideally unachievable by a uniformly distributed particle source, the present study uses experimental measurements obtained by the PAMELA spectrometer over the kinetic energy interval between 0.14 and 52 GeV~\cite{adriani2011pamela, adriani2013measurements}. The primary alpha particle energies are randomly sampled according to the corresponding spectral probability distribution and directed towards the lunar surface. 

The objective of this study is to analyze the contribution of the cosmic alpha particles to the generation of lunar albedo neutrons and gamma rays by using the GEANT4 simulations~\cite{agostinelli2003geant4}. A particular emphasis is placed on the energy distributions of secondary particles emerging from the lunar surface and the depth dependence of the neutron and gamma generation within the regolith. The simulation results are then utilized for the further characterization of alpha-induced secondary radiation and for an improvement in the understanding of cosmic-ray interactions with the lunar surface. This study is organized as follows. In section~\ref{Simulation_scheme}, the simulation scheme in the GEAN4 framework is described. While the simulations results are exhibited in section~\ref{Simulation_outcomes}, the conclusions are drawn in section~\ref{Conclusion}.
\section{Simulation scheme}
\label{Simulation_scheme}
\begin{figure}[H]
\begin{center}
\includegraphics[width=12cm]{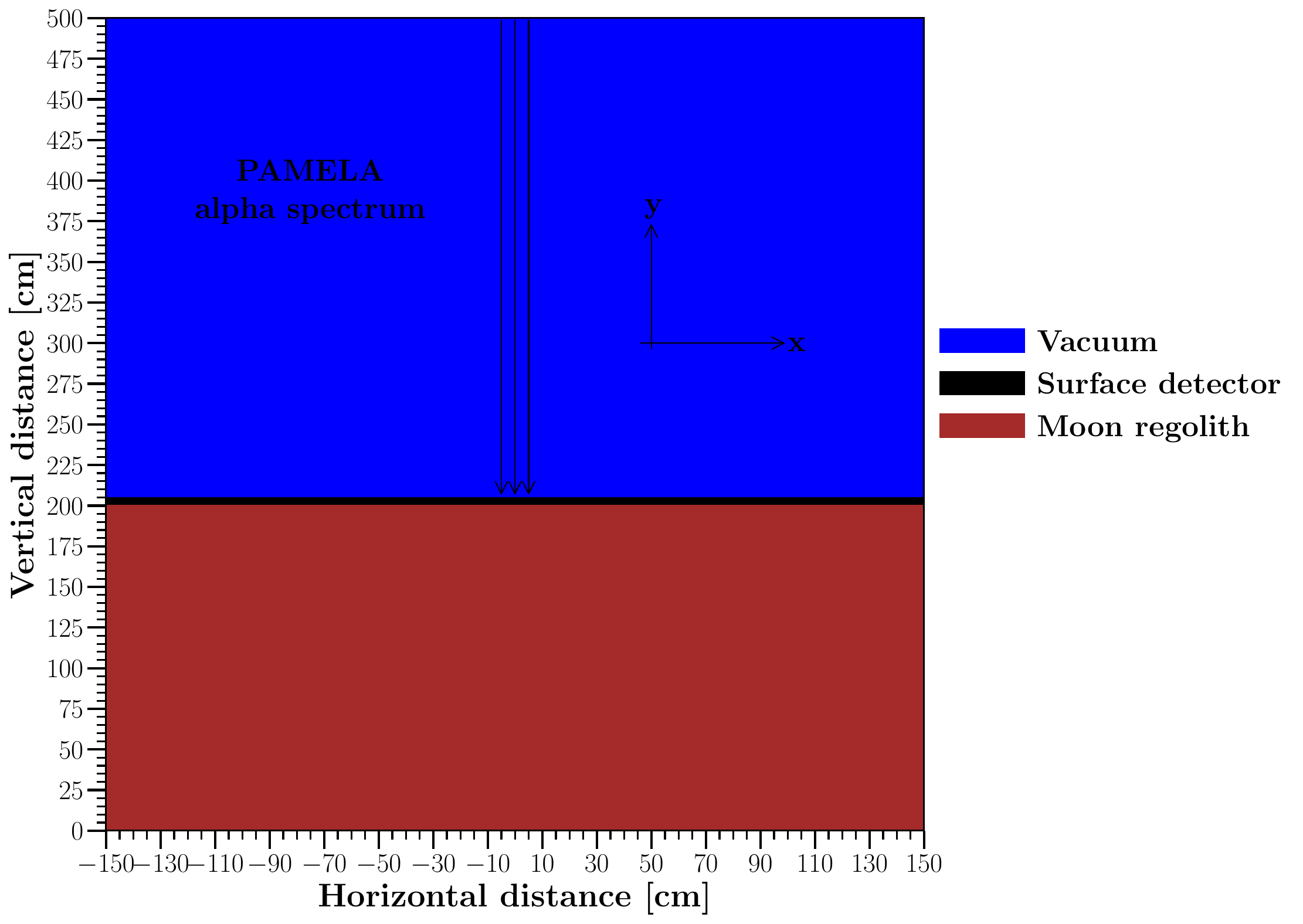}
\caption{Simulation setup for irradiation of lunar regolith with cosmic alpha particles in GEANT4.}
\label{Simulation_layout}
\end{center}
\end{figure}

\begin{figure}[H]
\begin{center}
\includegraphics[width=8cm]{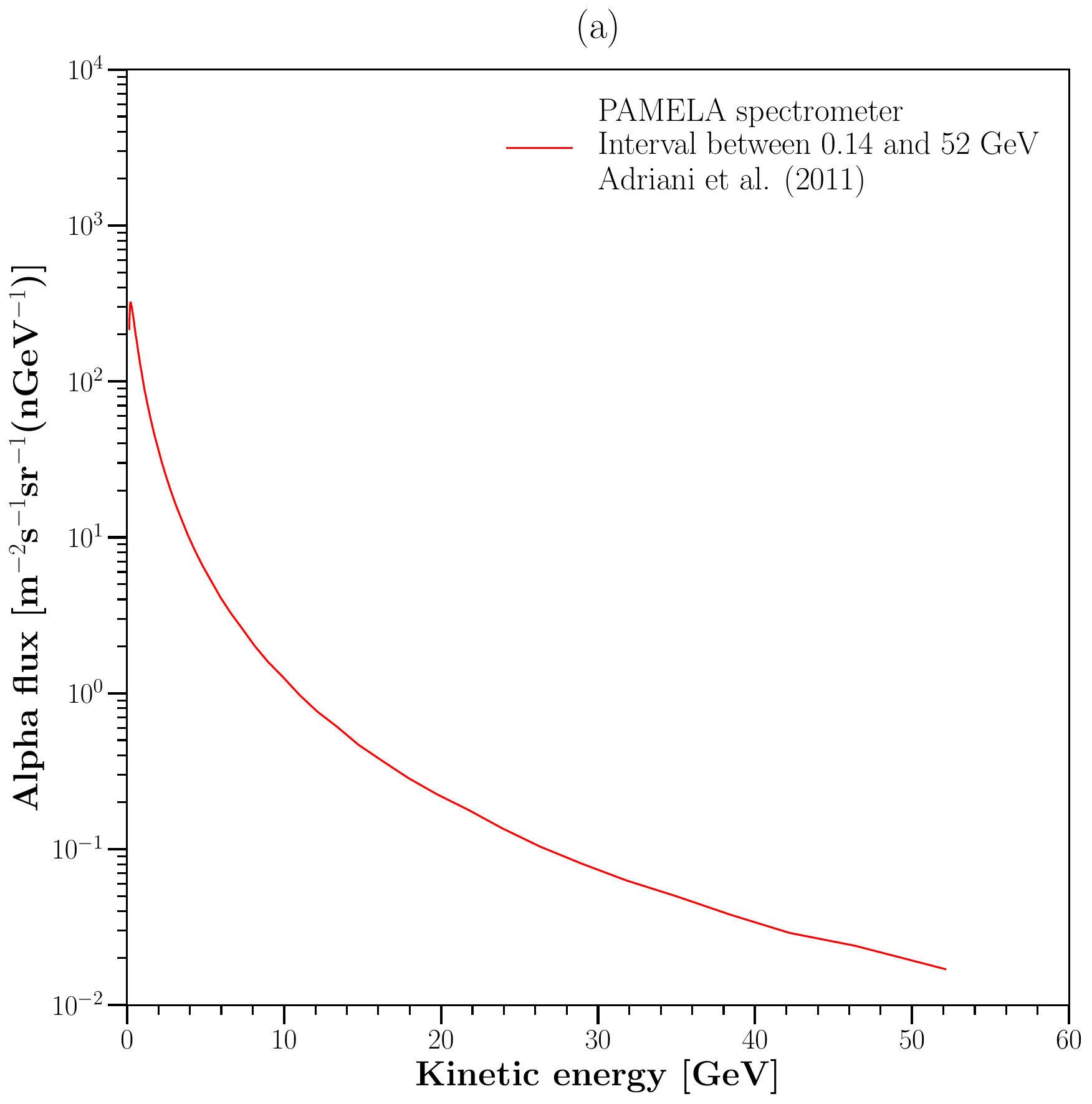}
\includegraphics[width=8cm]{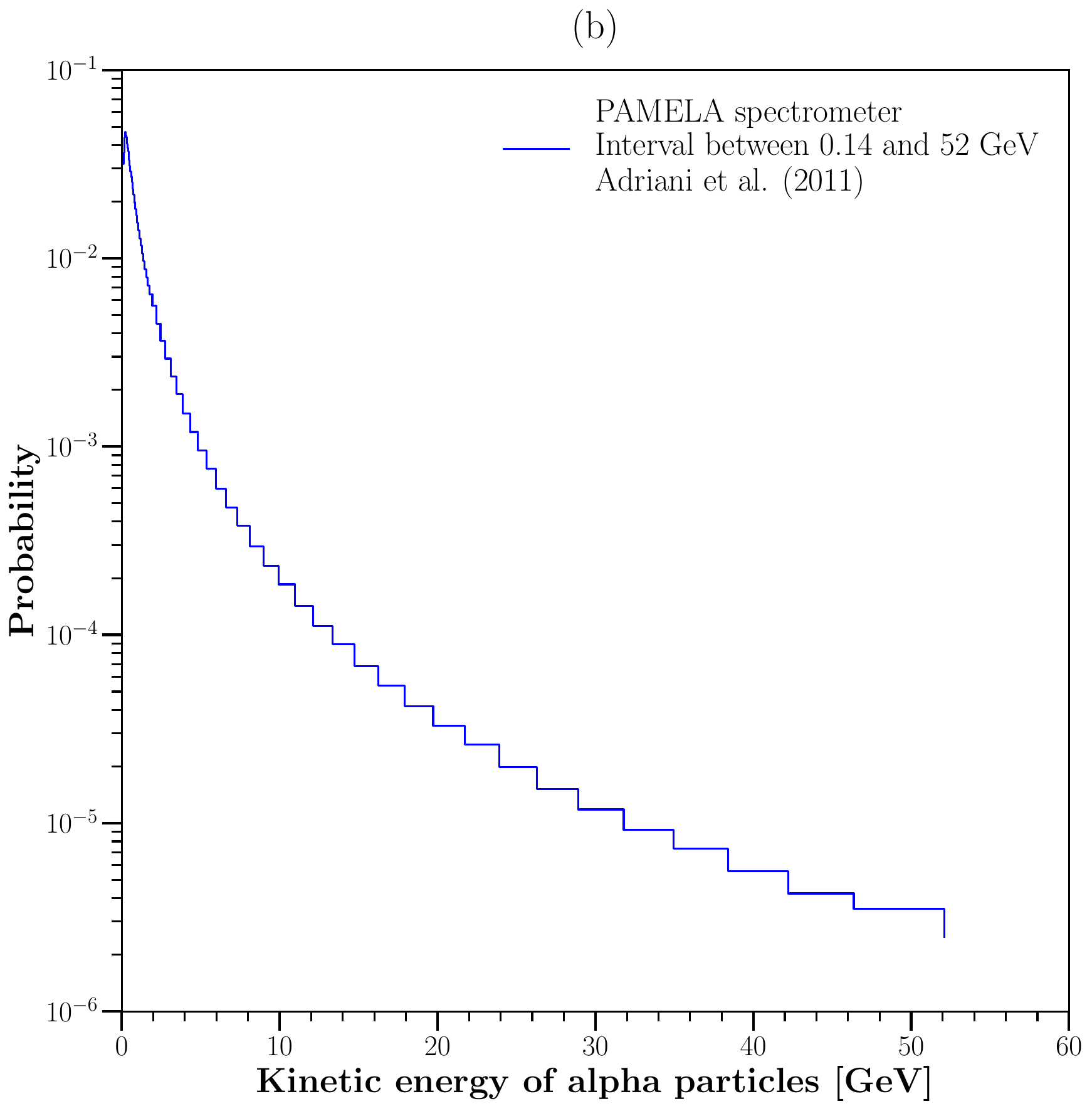}
\caption{Energy spectrum of cosmic alpha particles from the PAMELA spectrometer: (a) flux and (b) discrete probabilities.}
\label{Energy_spectrum_alpha}
\end{center}
\end{figure}

The Monte Carlo simulations based on the GEANT4 toolkit \cite{agostinelli2003geant4} are carried out to examine the generation of the secondary neutrons and gamma rays from the irradiation of lunar regolith by the cosmic alpha particles. As illustrated in Fig.~\ref{Simulation_layout}, the computational geometry consists of a lunar regolith volume of \ 3$ \times $2$ \times $3 m$^3$ that is surrounded by a vacuum environment providing a regolith depth of 200 cm, along with a density of 1.93 g/cm$^{3}$, following the previously developed voxelized bulk lunar surface geometry~\cite{TopuzGithubBulkLunar} and implemented in the dedicated simulation package released for the present study~\cite{TopuzGithubAlphaSurface}. A thin surface detector is positioned immediately above the lunar regolith as a scoring volume to record the secondary particles leaking out the lunar surface.
Nine elemental constituents are integrated according to their mass fraction as listed in Table~\ref{Table_composition}. This material composition consists of 43.7 wt.$\%$ oxygen, 0.3 wt.$\%$ sodium, 5.6 wt.$\%$ magnesium, 9 wt.$\%$ aluminum, 21.1 wt.$\%$ silicon, 8.5 wt.$\%$ calcium, 1.5 wt.$\%$ titanium, 0.1 wt.$\%$ manganese, and 10.2 wt.$\%$ iron, which is extracted from another study~\cite{zaman2022modeling}
\begin{table}[H]
\centering
\caption{Elemental composition of lunar regolith.}
\begin{footnotesize}
\begin{tabular}{c c}
\toprule
\toprule
Element & Mass fraction (\%) \\
\midrule
Oxygen (O) &\ 43.7 \\
Silicon (Si) &\ 21.1 \\
Iron (Fe) &\ 10.2 \\
Aluminium (Al) &\ 9.0 \\
Calcium (Ca) &\ 8.5 \\
Magnesium (Mg) &\ 5.6 \\
Titanium (Ti) &\ 1.5 \\
Sodium (Na) &\ 0.3 \\
Manganese (Mn) &\ 0.1 \\
\bottomrule
\bottomrule
\end{tabular}
\end{footnotesize}
\label{Table_composition}
\end{table}
The primary incident cosmic-ray particles are selected as the cosmic alpha particles that are directed vertically towards the lunar regolith along the negative (y)-axis. The primary energies of these particles are taken from the PAMELA spectrometer covering the kinetic-energy interval from approximately\ 0.14 to 52 GeV \cite{adriani2011pamela,adriani2013measurements}. The alpha flux decreases strongly with increasing kinetic energy over the selected interval as shown in Fig.~\ref{Energy_spectrum_alpha}(a).
To carry out the Monte Carlo simulations, The PAMELA alpha spectrum is converted into a discrete probability distribution with 68 discrete intervals as shown in Fig.~\ref{Energy_spectrum_alpha}(b), each associated with a probability derived from the measured spectral distribution, which follows the energy-discretization based particle-generation approach described in~\cite{topuz2022dome,ilker2023particle}. The kinetic energies are selected randomly according to the implemented probability distribution for each generated particle. The initial planar coordinates of the alpha particles are also randomly distributed over the irradiation region. The FTFP\_BERT\_HP reference physics list is utilized to model the particle interactions with matter and the transport. This physics configuration is applicable over a broad energy range and incorporates the high precision of the low-energy neutron transport. The simulation properties are summarized in Table~\ref{Simulation_properties}.
\begin{table}[H]
\begin{center}
\caption{Simulation properties.}
\begin{footnotesize}
\begin{tabular}{cc}
\toprule
\toprule
Particle & $\alpha$\\
Beam direction & Vertical\\
Momentum direction & (0, -1, 0)\\
Source geometry & Planar\\
Initial position (cm) & ([-0.5, 0.5], 85, [-0.5, 0.5])\\
Number of particles & 10$^{5}$\\
Energy interval (GeV) & [0.14, 52]\\
Energy distribution & PAMELA\\
Material database & G4/NIST\\
Reference physics list & FTFP$\_$BERT$\_$HP\\
\bottomrule
\bottomrule
\label{Simulation_properties}
\end{tabular}
\end{footnotesize}
\end{center}
\end{table}
A thin scoring volume is placed just after the lunar surface to identify the emerging secondary particles from the regolith material. During the particle transport, the GEANT4 stepping action is implemented to examine the properties of particles entering the surface detector. The records of each neutron and gamma ray are noted separately. For each detected particle, their parameters such as the track identification number, the spatial coordinates, the kinetic energy, the particle type, the detector volume, and the transport defining process step are stored.

The GEANT4 simulation outcomes are further analyzed by using Python. The neutron and gamma ray data files are imported to extract the kinetic energy values of the recorded particles. The particle energies are then aggregated into intervals to represent using histograms, and the corresponding energy distributions are obtained. The produced datasets are then further utilized to obtain the albedo neutron spectrum, the albedo gamma ray spectrum, their neutron generation depth, and the alpha range.
\section{Simulation outcomes}
\label{Simulation_outcomes}
\begin{figure}[H]
\begin{center}
\includegraphics[width=9cm]{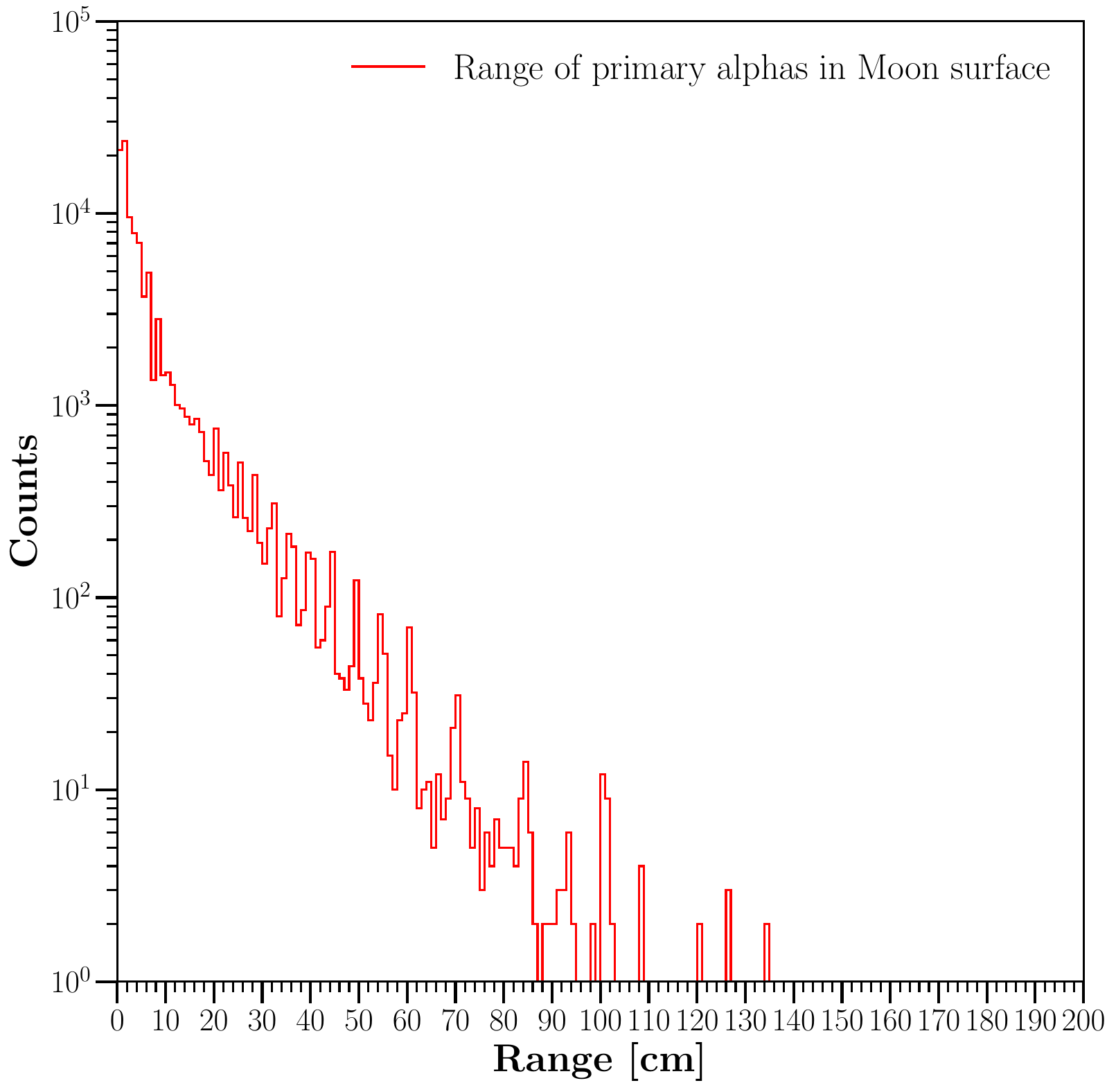}
\caption{Range of cosmic alpha particles with the lunar regolith.}
\label{Range_alpha}
\end{center}
\end{figure}
In the initial analysis, the penetration range of cosmic alpha particles into the lunar regolith is studied as demonstrated in Fig.~\ref{Range_alpha}. As the penetration depth increases, the range of the cosmic alpha particles is decreased, which also means that large amount of the incident alpha particles travel very short range within the lunar regolith. As a result, the number of particles reaching depths decreases. The simulation setup shows the alpha particle range distribution up to approximately 200 cm. This behavior can be associated with the broad range incident alpha particles referred from the PAMELA-based spectrum over the kinetic-energy range between 0.14 and 52 GeV. The incident alpha particles with different energies exhibit different penetration ranges, and a low number of alpha particles propagate into the deeper regions of regolith. These penetration patterns are relevant to the depth-dependent production of the secondary particles.
\begin{figure}[H]
\begin{center}
\includegraphics[width=8cm]{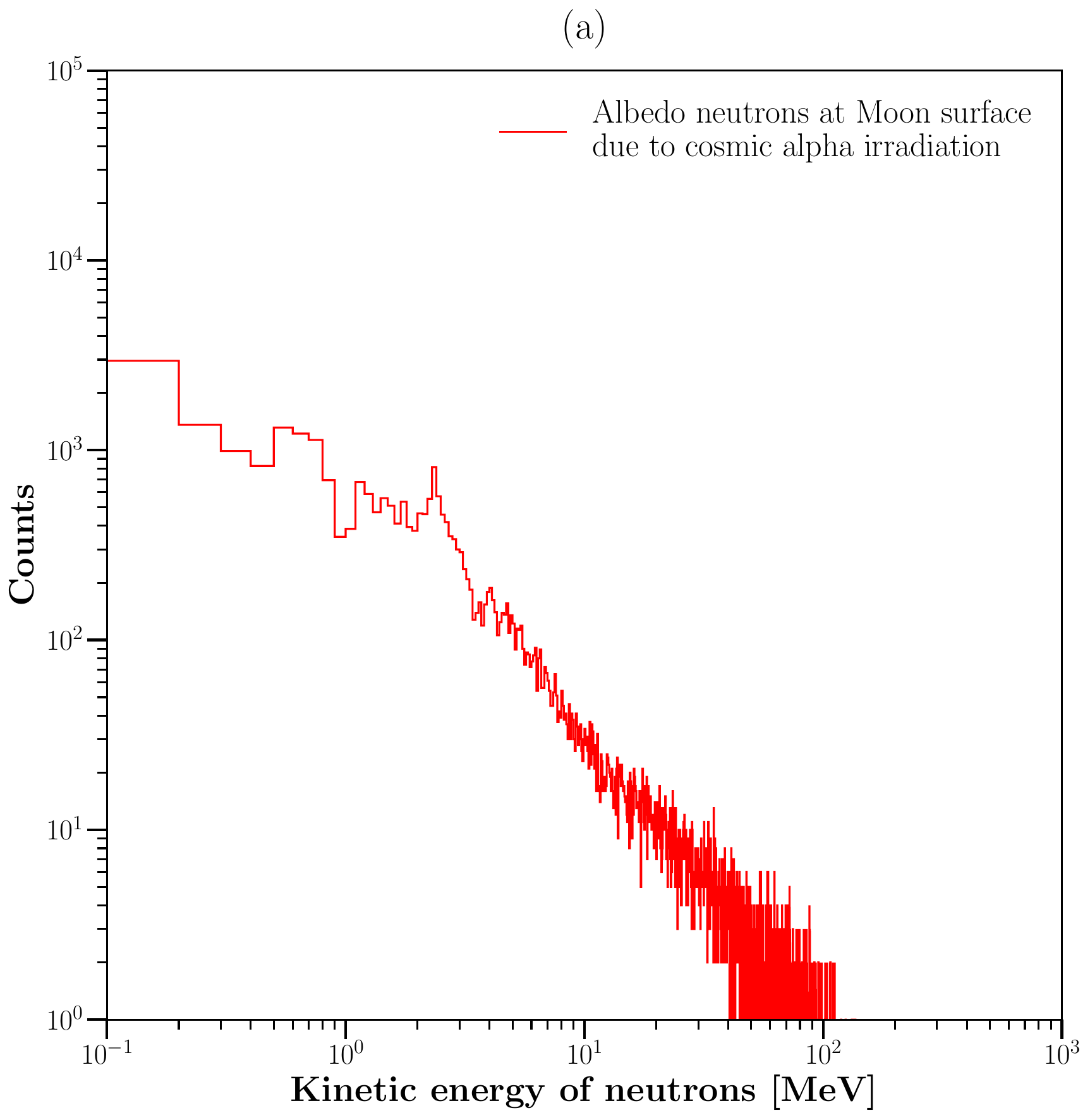}
\includegraphics[width=8cm]{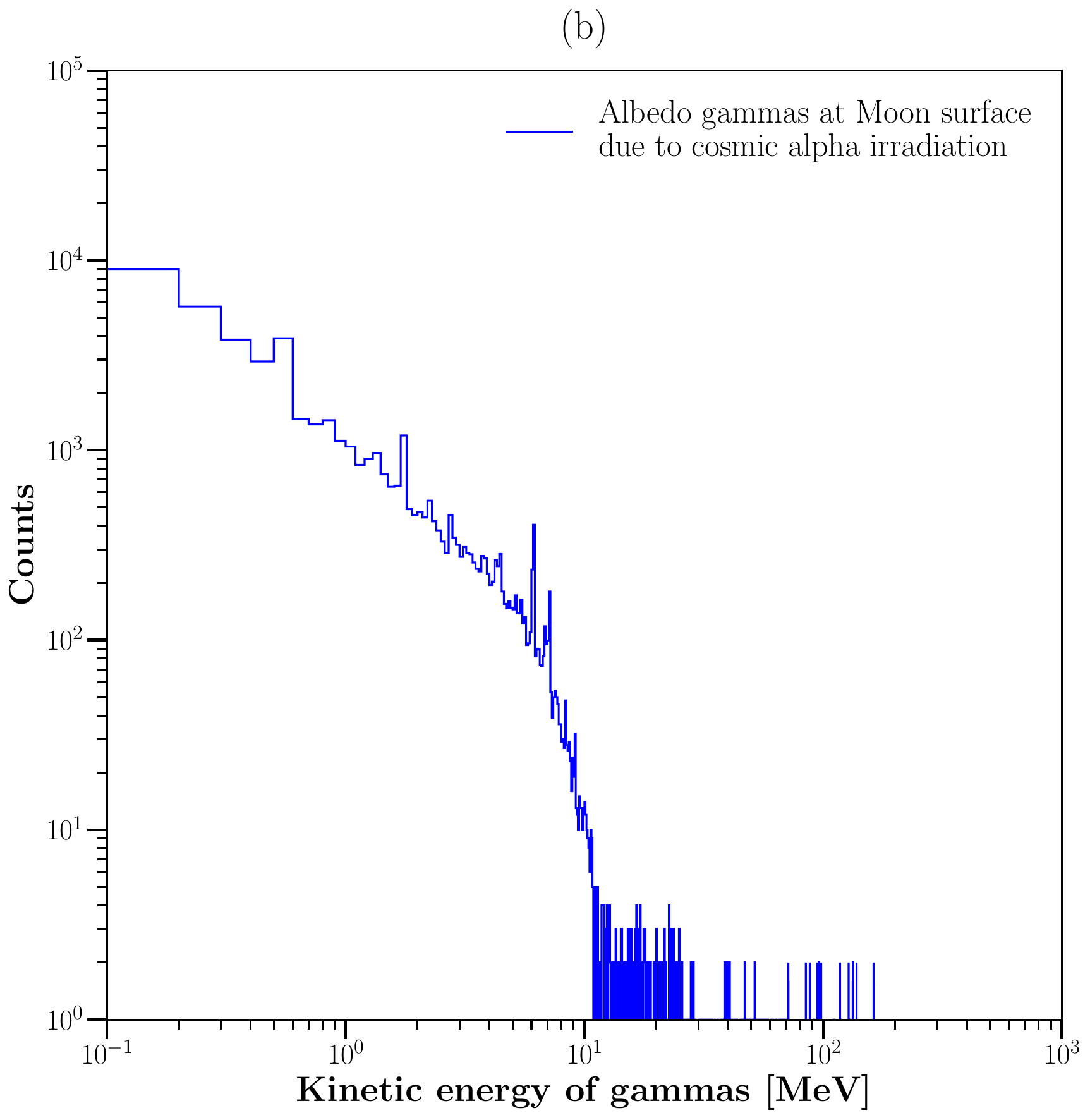}
\caption{Energy spectrum of secondary radiation due to bombardment of lunar surface with cosmic alpha particles: (a) albedo neutrons and (b) albedo gammas}
\label{Energy_spectrum_albedo_neutron_gamma}
\end{center}
\end{figure}
In the next stage, the generated secondary neutrons escaping from the lunar regolith are recorded by a pseudo surface detector positioned above the lunar surface. Correspondingly, the resulting energy spectrum of the generated albedo neutrons is shown in Fig.~\ref{Energy_spectrum_albedo_neutron_gamma}(a). The albedo neutron spectrum extends over a broad kinetic-energy interval and depicts a decreasing pattern on the logarithmic scale. The largest neutron population is observed over the low-energy region, and a significant decrease is observed in the higher kinetic-energy region. Even though the spectrum extends towards the order of MeV, few neutrons are observed in the high-energy region. This spectral behavior is a result of both the initial production and the subsequent transport of albedo neutrons through the regolith. The alpha-induced neutrons may undergo through different nuclear reactions and scattering before they escape from the surface. This transport phenomenon results in the modification of the energy distribution of the albedo neutrons and have larger population at comparatively lower energies. The high energy component depicts that only a small number of albedo neutrons with the high kinetic energies reach to the lunar surface. Thus, it demonstrates that albedo neutron population generated by the cosmic alpha irradiation is predominantly shifted towards the lower-energy region, with a gradual decrease in population towards higher energies.

\begin{figure}[H]
\begin{center}
\includegraphics[width=7.8cm]{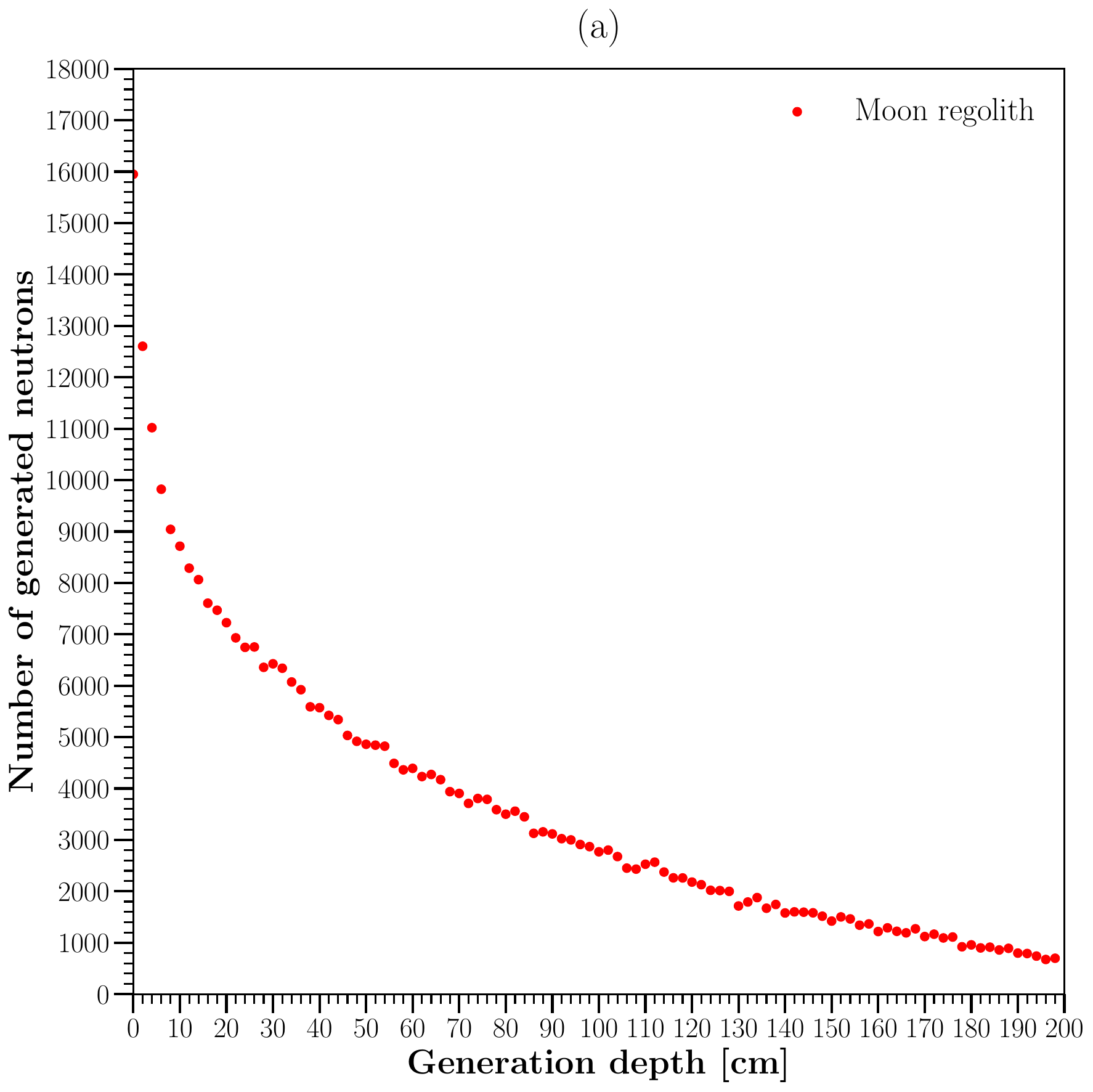}
\includegraphics[width=7.8cm]{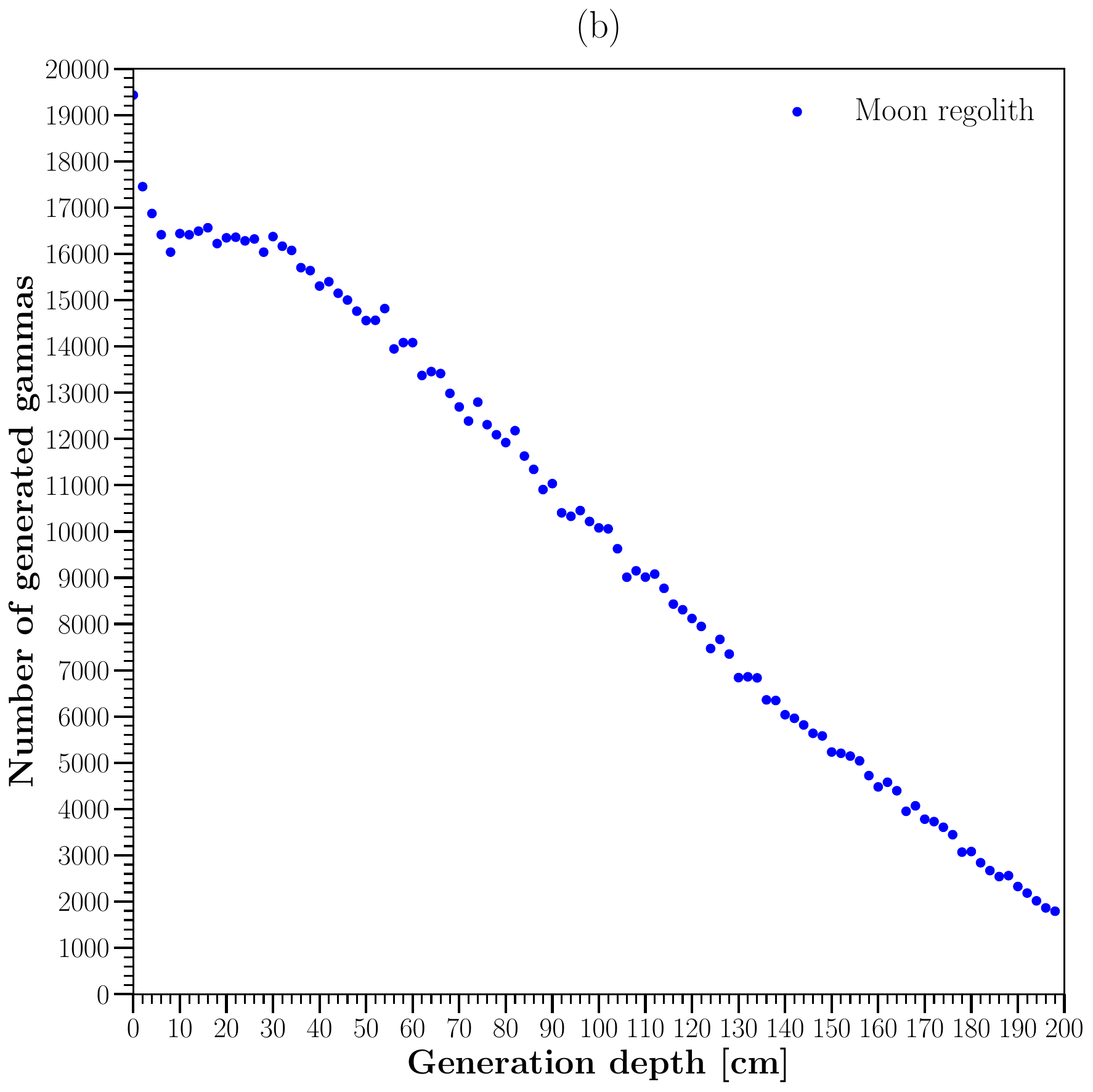}
\caption{Generation depth of secondary particle in lunar regolith due to cosmic alpha irradiation: (a) secondary neutrons and (b) secondary gammas}
\label{Generation_depth}
\end{center}
\end{figure}

The generated secondary gamma rays escaping the surface are similarly recorded by using the pseudo surface detector, and the resulting gamma-ray energy is depicted in Fig.~\ref{Energy_spectrum_albedo_neutron_gamma}(b). The simulated gamma-ray spectrum also exhibits a strong energy-dependent behavior. Comparatively, a large number of gamma ray population is observed at the lower energies, followed by a significant decrease in the counts with the ascending gamma ray energy. A quantifiable reduction in the population is visible in several MeV regions, while a sparse high energy component extend towards and beyond the MeV region. The albedo gammas originate from the nuclear processes occurring during the alpha-induced particle cascade within the lunar regolith. The incident primary alpha interactions produce the secondary albedo neutrons which subsequently interact with the constituent nuclei of the lunar regolith material. The nuclear interactions and de-excitation process are carried out by the particles throughout this cascade resulting into the production of the gamma rays. The low-energy gamma rays are majorly observed escaping the lunar surface. However, the presence of high-energy tails demonstrates that a part of produced radiation includes energetic gamma rays as a result of alpha-induced nuclear interactions and subsequent secondary-particle process.
\begin{figure}[H]
\begin{center}
\includegraphics[width=8cm]{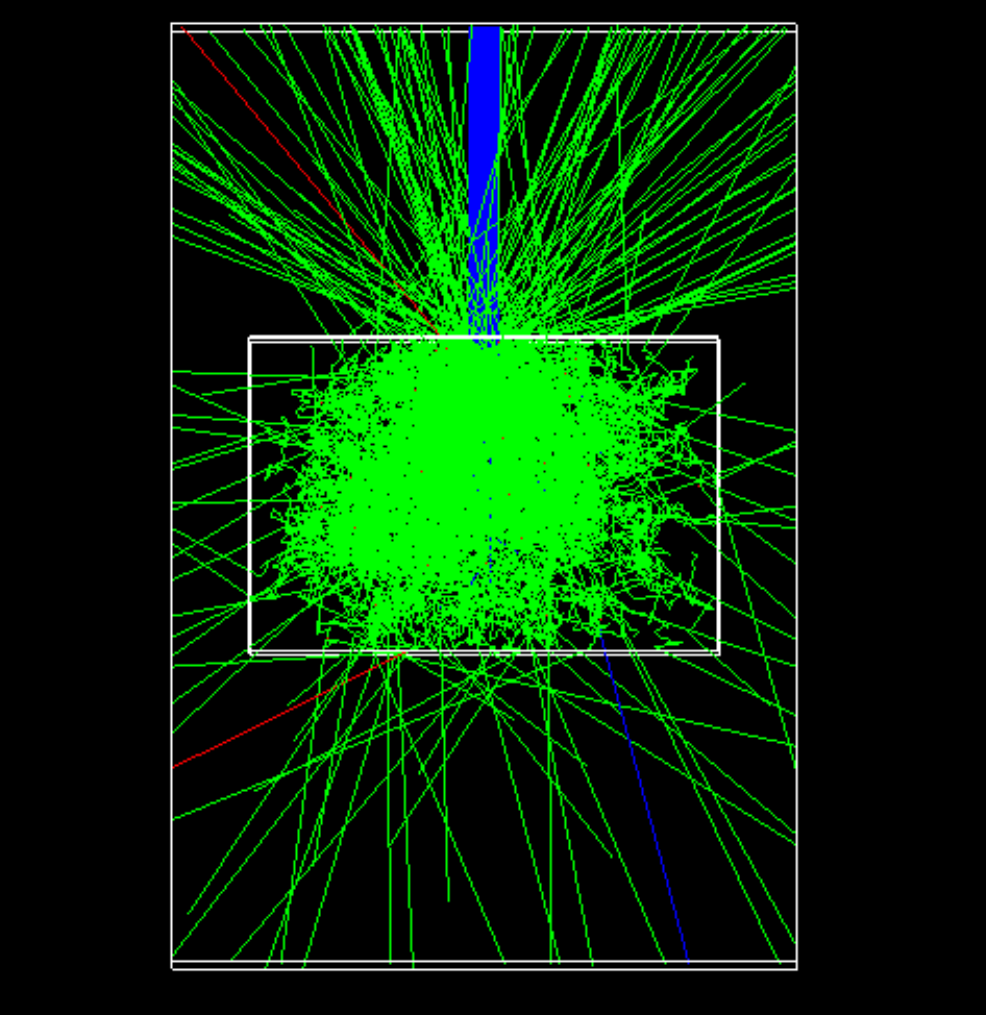}
\caption{Secondary neutral particle emission (green) from lunar regolith due to alpha irradiation (blue).}
\label{Particle_emission}
\end{center}
\end{figure}
As displayed in Fig.~\ref{Particle_emission}, the integrated simulation results indicate that the cosmic alpha particles contribute to the production of both secondary neutrons and gamma rays in the lunar regolith. The generation depth curve in Fig.~\ref{Generation_depth} shows that most of the secondary particle generation is concentrated primarily within the upper regolith region and decrease towards the greater depths. 
The evaluated surface spectra further demonstrate that both albedo neutrons and gamma ray populations are primarily governed by the lower-energy particles, with gradually diminishing counts at higher values. Regardless of the smaller fraction of alpha particles in the galactic cosmic rays compared with the protons, their interaction with the multiple elements present in the lunar regolith initiates nuclear cascades capable of producing detectable secondary radiation. As a result, the present simulation isolates the contribution of cosmic alpha component and demonstrates its role in the generation of the lunar albedo neutrons and gamma ray environment, complementing earlier findings on the proton- and hydrogen-related contributions to the lunar albedo neutron population \cite{topuz2024effect} as well as global dose-rate assessments of the lunar radiation environment \cite{naito2023global}.
\section{Conclusion}
\label{Conclusion}
The current computational analysis investigates the contribution of the cosmic alpha particles to the generation of the lunar albedo neutrons and gamma rays in the lunar regolith by using the GEANT4 simulations. The PAMELA alpha particle spectrum within the energy range of 0.14 to 52 GeV is taken into consideration for the incident primary particles. The derived results show that the population of the alpha particles gradually decreases with the increasing penetration range. The generation of the secondary particles due to the alpha irradiation is mostly concentrated in the near-surface region and shows a quantifiable decrease at the greater regolith depth. The obtained energy spectra also depict a higher distribution of secondary particles in the lower-energy regions, followed by a significant drop towards the higher energies. Therefore, the results indicate that the cosmic alpha particles, although they constitute a smaller fraction than the protons in the cosmic rays, yet make a prominent contribution to the lunar radiation environment. This study also demonstrates the utility of GEANT4 simulations in understanding cosmic-ray interactions and secondary particle production within the lunar regolith.
\bibliographystyle{elsarticle-num}
\bibliography{Moon_alpha.bib} 
\end{document}